*Graphical Table of Contents*

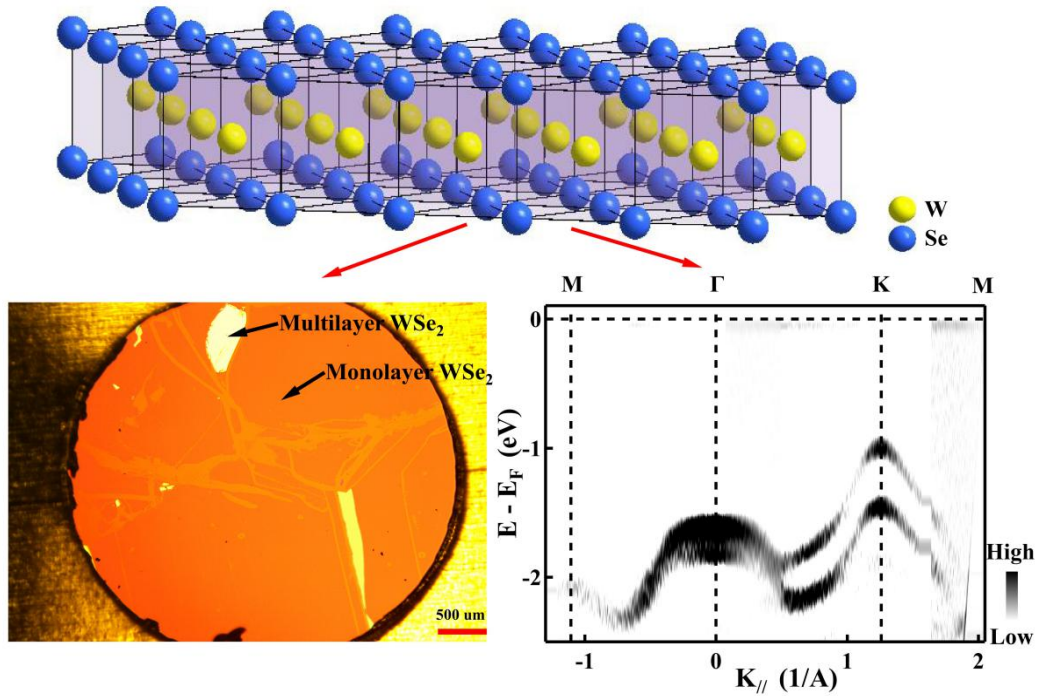

The monolayer WSe$_2$ is interesting and important for future application in nanoelectronics, spintronics and valleytronics devices. We have successfully fabricated a few millimeter-sized monolayer WSe$_2$ single crystals and carried out comprehensive electronic band structure study on such high quality samples, using standard high resolution angle-resolved photoemission spectroscopy.

# Electronic Structure of Exfoliated Millimeter-Sized Monolayer WSe$_2$ on Silicon Wafer

Wenjuan Zhao[1,2], Yuan Huang[1,2*], Cheng Shen[1,2], Cong Li[1,2], Yongqing Cai[1,2], Yu Xu[1,2], Hongtao Rong[1,2], Qiang Gao[1,2], Yang Wang[1,2], Lin Zhao[1], Lihong Bao[1], Qingyan Wang[1], Guangyu Zhang[1], Hongjun Gao [1,2], Zuyan Xu[3], Xingjiang Zhou[1,2,4,5*], Guodong Liu[1,4*]

[1] Beijing National Laboratory for Condensed Matter Physics, Institute of Physics, Chinese Academy of Sciences, Beijing 100190, China

[2] University of Chinese Academy of Sciences, Beijing 100049, China

[3] Technical Institute of Physics and Chemistry, Chinese Academy of Sciences, Beijing 100190, China

[4] Songshan Lake Materials Laboratory, Dongguan, Guangdong 523808, China

[5] Beijing Academy of Quantum Information Sciences, Beijing 100193, China

## Corresponding Authors

*E-mail: yhuang01@iphy.ac.cn
*E-mail: xjzhou@iphy.ac.cn
*E-mail: gdliu_arpes@iphy.ac.cn

# Abstract

The monolayer WSe$_2$ is interesting and important for future application in nanoelectronics, spintronics and valleytronics devices, because it has the largest spin splitting and longest valley coherence time among all the known monolayer transition-metal dichalcogenides (TMDs). To obtain the large-area monolayer TMDs' crystal is the first step to manufacture scalable and high-performance electronic devices. In this letter, we have successfully fabricated millimeter-sized monolayer WSe$_2$ single crystals with very high quality, based on our improved mechanical exfoliation method. With such superior samples, using standard high resolution angle-resolved photoemission spectroscopy, we did comprehensive electronic band structure measurements on our monolayer WSe$_2$. The overall band features point it to be a 1.2eV direct band gap semiconductor. Its spin-splitting of the valence band at *K* point is found as 460 meV, which is 30 meV less than the corresponding band splitting in its bulk counterpart. The effective hole masses of valence bands are determined as 2.344 m$_e$ at *Γ*, and 0.529 m$_e$ as well as 0.532 m$_e$ at *K* for the upper and lower branch of splitting bands, respectively. And screening effect from substrate is shown to substantially impact on the electronic properties. Our results provide important insights into band structure engineering in monolayer TMDs. Our monolayer WSe$_2$ crystals may constitute a valuable device platform.

**Keywords:** *transition-metal dichalcogenides, WSe$_2$, monolayer, electronic structure, angle-resolved photoemission spectroscopy*

# 1 Introduction

Two-dimensional (2D) crystals, such as graphene[1], hBN[2], phosphorene[3], transition metal dichalcogenides (TMDs) [4,5] and magnetic layered compounds [6, 7] etc., constitute a unique platform for novel physical properties and functionalities not existing in their bulk counterparts. Particular focus has been on the TMDs $MX_2$ (M = Mo, W; X = S, Se, Te) semiconductors in the past decade, a family of materials that have very distinct properties in their monolayer limit, such as indirect to direct band gap transition from few layer to monolayer[8-12], obvious spin-splitting of the valence band [10,13], well-defined valley degrees of freedom and spin-valley locking [14-16], spin-layer locking [17], huge exciton binding energy [18-21], etc. Based on these superior properties, many application related studies have been made for functional testing and device demonstration in the field of photoluminescence, laser, field effect transistors (FETs)[22,23], photodetector[24], and so on. The 2D atomic crystals of TMDs have shown great potential in the next-generation electronics, optoelectronics[25] and spintronics [26].

Among all the TMDs semiconductors, $WSe_2$ is a special and important family member for basic research and potential applications. Besides having a direct band gap which is in favor of electronic and optoelectronic applications, the monolayer $WSe_2$ possesses the largest spin-splitting in the valence bands near *K/K′* of Brillouin zone[27], which is relevant to AB exciton effect, spin-valley coupling and spin layer coupling. Such a large spin-splitting of monolayer $WSe_2$ would be advantageous to fabricate more effective spintronic devices. The $WSe_2$ monolayer therefore becomes an ideal target for investigating spin and valley dependent properties as well as for optoelectronic and spintronic applications. Furthermore, the other advantage of monolayer $WSe_2$ is that it preserves a longer valley coherence time than other TMDs [26,28], which makes it a more promising material for valleytronics. Since the low energy electronic structure governs the base of physical properties and applications in most materials, it is particularly important to experimentally establish the spin-valley locked band structure of monolayer $WSe_2$. This can greatly help to design optoelectronic and spintronic devices on the basis of the band-structure engineering. Up to now, there have been few works on electronic structure measurements of monolayer $WSe_2$, with one for molecular beam epitaxy (MBE) grown film [27] and others for exfoliated micro flake[29,30]. In the case of former, the monolayer film contains triangular-shaped twin domains, that is not in favor of scalable device manufacturing. And no experimental band along *Γ-M* and valence band mapping are reported. In the later case, limited by sample size and instrument resolution, the data quality is rather poor. And in those measurements, it lacks data of the clear constant energy contour of valence bands in ref. 29 and 30 and the critical bands along high symmetry directions *Γ-M* and *K-M* in ref. 30. A comprehensive electronic structure study for the large-area and high-quality $WSe_2$ monolayer is highly desirable in photoemission experiments.

In this letter, by using a modified mechanical exfoliation method [31], we have succeeded in fabricating millimeter-sized monolayer single crystal $WSe_2$ sample with high quality. Using high resolution angle-resolved photoemission spectroscopy (ARPES), we report the systematic low-energy electronic structure studies for monolayer and bulk $WSe_2$. Our study provides comprehensive and clear electronic structure information of exfoliated and large-area monolayer $WSe_2$. We also present the first direct experimental evidence of quantum well state formation after Rubidium (Rb) doping. Our results provide much helpful information for understanding the exotic physical properties in monolayer $WSe_2$ and developing scalable and high performance electronic devices based on TMDs materials.

## 2  Experimental

We prepared the monolayer $WSe_2$ samples from bulk crystal by a modified mechanical exfoliation method [31]. The high quality 2H-$WSe_2$ single crystals (HQ Graphene Inc.) were grown by the chemical vapor transport method. First, the $SiO_2$/Si substrate was cleaned by oxygen plasma to desorb any organic molecules. Second, a ultra thin Au layer (~ 20nm) was evaporated onto the substrate, in order to reliably ground $WSe_2$ sample in ARPES measurements. Then, a thin $WSe_2$ crystal layer attached on the Scotch tape was pressed onto the Au-capped substrate to leave a monolayer on its surface. With such a procedure, we successfully exfoliated the millimeter-scale monolayer $WSe_2$ samples onto the substrate, as demonstrated in Fig.1c and 1d for optical image and Raman spectra characterization, respectively. The ARPES measurements on both monolayer and bulk $WSe_2$ samples were carried out at low temperature of about 30K by using our home-build Photoemission spectroscopy system with a VUV5000 Helium lamp and a DA30L electron energy analyzer of Scienta Omicron[32]. The He I$\alpha$ resonance line (21.218 eV) was used to excite photoelectrons. The energy and angular resolutions were set at 10-20 meV and 0.2 degree, respectively. During the whole measurements, the ultrahigh vacuum was kept at better than $5 \times 10^{-11}$ mbar.

## 3  Results and Discussion:

Fig.1 shows characterization of the millimeter-sized $WSe_2$ monolayer sample. In the monolayer 1H-$WSe_2$, it also presents a sandwich Se-W-Se layer with a trigonal prism unit cell (Fig.1a). W and Se atoms with different position along c axis form a honeycomb lattice structure (Fig.1b). The adjacent layers in bulk $WSe_2$ point to opposite direction in the *ab* parallel plane. Among all the $MX_2$, $WSe_2$ has the lattice parameters of a=3.28 Å, c=12.98 Å[33], which is similar to many other $MX_2$ crystals.

The optical image of a typical $WSe_2$ monolayer sample is displayed in Fig. 1c. Judging from the optical intensity contrast [34, 35], the most light gray area is corresponding to a monolayer with a uniform mirror-like surface. The small dark and white areas represent few layer and substrate, respectively. The thickness of our monolayer sample was further confirmed by AFM measurement to be ~ 6.5 Å, which is matching well with the reported lattice parameter [36,37]. For 2D materials, Raman spectroscopy is believed to be a convenient tool in determining layer thickness and uniformity [38]. Raman spectra were taken at a series of spots on the monolayer $WSe_2$ sample, as marked by red diamonds in Fig.1c. These spectra in Fig.1d reveal the characteristic and consistent $A_{1g}$ mode (251cm$^{-1}$) peaks[39], with the almost same peak position and intensity crossing millimeter-sized area of $WSe_2$ monolayer. It provides a strong evidence of the good homogeneity and high quality of our monolayer WSe$_2$ sample. The advantage of our exfoliated large area samples is that they are real single crystal and lacking for grain boundaries and defects, compared with those grown by CVD or MBE methods. Such exfoliated samples are of great benefit to both intrinsic property study and high-performance device production.

Fig. 2 shows the measured electronic structure of our monolayer WSe$_2$. Fig. 2a-f present the constant energy contour evolution from small circle, to single triangle and to concentric double triangles, with the increased binding energy. Such concentric triangular warping is consistent with the anisotropic band splitting around *K/K'* point. It reflects the anisotropy of electronic states around *K/K'*. The triangular warping of photoemission intensity agrees with the three-fold rotational symmetry of the crystal structure, which can be simply approximated as third order correction to the parabolic energy dispersion [41].

Fig. 2h displays the low-energy band structure along the *M-Γ-K-M* high symmetry direction. To reduce background and highlight the intrinsic band signals, Fig. 2i illustrates the second-derivative spectra of original data in Fig.2h with respect to energy. The corresponding band calculation data in ref. [10,40] are also shown in Fig. 2i as red and blue circles, after an appropriate energy shift for adjusting chemical potential. We can find excellent agreement between the measured band dispersion and calculated band structure with spin orbit coupling included.

The near-E$_F$ valence band of monolayer WSe$_2$, as displayed in Fig. 2h-i, features a single flat band around *Γ* and a large band splitting in *K* region. From *M* to *Γ* to *K* and back to *M*, one single spin-degenerate band starts to split into two spin-resolved bands at one third position of $\Gamma$-$K$ line. These features are actually common in all the 1H monolayer MX$_2$ [13,41,42]. Furthermore, the valence band maximum (VBM) at $\Gamma$ is located at rather higher binding energy (-1.7 eV) than that (-1.02 eV) at $K$. Different from the bulk WSe$_2$, here the position of global VBM is switched from $\Gamma$ into $K$.

From the density functional theory (DFT) calculations [43,44], it can be identified that the flat band around $\Gamma$ is formed by most W $d_{z2}$ orbital and little Se p$_z$ orbital. In WSe$_2$ lattice, these orbitals are extending out of plane but localized in plane, resulting in full flatness of this band. While two splitting valence bands around $K$ originate mainly from a combination of W $d_{x2-y2}$ and $d_{xy}$ orbitals, which are rather itinerant in plane. This leads to a strong dispersion of these electronic states. And spin orbital coupling arising from W atoms induce a large up-and-down splitting of the $K$-region bands.

In monolayer MX$_2$ family, many intriguing physical phenomena and properties are governed by the band splitting at Brillouin zone corner, such as indirect-direct gap transition, spin-valley locking, and AB exciton effect, etc. The band splitting at $K$ along $\Gamma$-$K$ direction is comparatively displayed in Fig. 3 for monolayer and bulk WSe$_2$ sample, respectively. The clear and sharp band dispersion can be observed, as shown in Fig. 3a, b. To facilitate our quantitative extraction of the specific splitting size, we take the energy distribution curves (EDCs) at $K$ point of original bands, which are shown in Fig. 3c and 3d (red and orange solid lines). From these EDCs, the splitting size can be derived as 460 meV in monolayer and 490 meV in bulk. Such a splitting increase with layer stacking is consistent with the findings in other MX$_2$ materials from photoluminescence measurement [45-47] and band calculations [14,48] reported recently.

Like other few 1H $MX_2$ monolayer, DFT band calculation predicts monolayer $WSe_2$ to be a 2D direct-gap semiconductor, that has been proven by ARPES measurement in the MBE-grown $WSe_2$ sample [27]. In order to directly determine the band gap character of our exfoliated $WSe_2$ monolayer, its surface was in situ doped by alkaline metal Rb to pull the conduction band below Fermi level. After adequate Rb doping at low temperature, as seen in Fig. 4d, the conduction band minimum (CBM) at *K* indeed appears near Fermi edge. In Fig. 4a and 4b along *Γ-K-M* direction, we compare the original band dispersion for pristine $WSe_2$ monolayer with that for Rb-doped sample. To quantitatively evaluate the band shifting, the EDCs at *Γ* and *K* are plotted in Fig. 4e and f, respectively. For the pristine sample, the VBM is located at *K* with a binding energy of -1.02 eV, but the conduction band is invisible. After doping by Rb (Fig. 4b), the whole valence band moves down to higher binding energy. Yet, no conduction band can be identified. While it can be seen clearly as an additional weak spectral weight patch near $E_F$, when we took a proper measurement geometry, as demonstrated in Fig 4d. That is, the spectra were acquired along the cut 2 as marked in Fig. 4c, under *s* polarization of photon. In this case, the dipole matrix element integral becomes finite, since the in-plane polarization component plays an important role, although the W $d_{z2}$ orbital-derived CBM has even parity. In other case as shown in Fig. 4a-b, the matrix element integral is equal to zero in principle, that is the reason why we cannot observe CBM at *K*. Considering the CBM being aligned together with the VBM with respect to momentum, it indicates that our monolayer WSe$_2$ is a direct gap semiconductor. The gap size at *K* is determined to be 1.2 eV in Fig.4d, that is close to the value of 1.40 eV measured by ARPES and much less than the value of 1.95 eV obtained from STS measurement in a MBE grown monolayer WSe$_2$ [27]. It is because the screening effect from the substrate gold layer reduces the long range coulomb interaction between electrons in WSe$_2$ monolayer.

With Rb doping, we found that the VBM at *K* shifts down by 0.31 eV, while band maximum at *Γ* moves down by 0.4 eV. Apparently the Rb-doping induced band shift is not a rigid shift in monolayer WSe$_2$. This is mostly because the bands around *Γ* and *K* have different orbital origin. The dominant orbital of valence band near *Γ* is the out of plane orbital $d_{z2}$ from W atoms, which is more susceptible to electron doping than the in-plane orbital $d_{x2-y2}/d_{xy}$ for the band around *K*. This surface doping strategy for monolayer WSe$_2$ as well as other monolayer TMDs provides a promising avenue of electronic structure engineering.

With heavy Rb doping, we could observe the quantum well state (QWS) confined in the Rb layer on monolayer WSe$_2$. Similar QWS was also investigated by Alidoust et al. on MoSe$_2$ single crystal surface after potassium deposition [49]. Fig.4g presents the corresponding band dispersion, with a *Γ*-centered and near-$E_F$ shallow band being QWS band. It is also guided by a red-dashed parabolic line. On its right side, we can see more spectral weight induced by populating the CBM at *T* point (a point approximately at midway in the *Γ* − *K* line). Besides these two near-$E_F$ bands, a non-dispersive band marked by a horizontal green line, appears at ~ 0.5eV binding energy. It sits on the same energy level as that in Rb metal, where a density-of-state extreme exists [50,51]. It is reasonable to attribute such a flat band to an impurity band from the scattered Rb cluster on monolayer $WSe_2$ surface. In Fig. 4h, the QWS-derived Fermi surface is schematically plotted as a small red circle. While the Fermi surface from CBMs at $T$ point forms a hexagonal color patch. Here, the formation of quantum well state can be specifically explained as coming from two-dimensional electron gas confined in epitaxial Rb layer, sandwiched in between two barrier layers of vacuum and monolayer $WSe_2$ semiconductor [49, 52]. Depending on the layer number of conductive Rb atom layer, there may be more QWSs appearing. Such QWSs on monolayer $WSe_2$ may provide us with an important knob to manipulate its electronic properties, which could be useful in application of future electronics and optoelectronics.

For monolayer $WSe_2$, our results represent the very complete and high-quality electronic structure dataset achieved by ARPES up to now. Zhang et al. [27] reported high-quality band structure data on the MBE grown monolayer $WSe_2$, but they did not make measurements along $\Gamma$-$M$ direction and even along $K$-$M$ direction. Even if they showed some data partially along $K$-$M$ direction, it was hard to resolve the bands due to interference from graphene bands. Duy Le and Wilson et al. presented band dispersions along $\Gamma$-$K$ and $M$-$K$ directions for the exfoliated monolayer $WSe_2$[29,30]. However, limited by sample size and deficient monolayer quality, it's hard to pin down some important band characteristics. Therefore, our present study provides comprehensive and clear electronic structure information of exfoliated and large-area monolayer WSe$_2$.

For monolayer MX$_2$, the band splitting around *K* is caused by intra-layer spin-orbit coupling and inversion symmetry breaking. Therein the two split bands become spin splitting, except that the bands along *Γ−M* line remain their spin degeneracy due to time reversibility [53]. Our results reflect clearly such character. Among all monolayer MX$_2$ semiconductors, monolayer WSe$_2$ has the largest spin splitting size (460 meV), which is much larger than splitting in MoSe$_2$ (~180 meV) [13] , MoS$_2$ (~145 meV) [42] , MoTe$_2$(~212 meV) and WS$_2$(~419 meV)[41]. This is because the spin–orbit coupling strength is enhanced by the heavy atoms W and Se. The monolayer WSe$_2$ thereby may be the most promising material in spintronics application among MX$_2$ family. It is noted that the measured band splitting energy in monolayer WSe$_2$ by ARPES was reported as 475 meV on bi-layer graphene [27] and 513 meV on a SiO$_2$/Si substrate [29]. It looks that screening or Rashba effect from substrate may have a profound effect on such a spin splitting energy. The specific role and mechanism in this regard deserve further study in future.

In bulk WSe$_2$, besides the intra-layer spin-orbit coupling (SOC), van der Waals interlayer coupling also contributes to the band splitting around *K*. Since the SOC between layers is negligible [48], the SOC induced band splitting size at *K* can be regarded as the same between monolayer and bulk WSe$_2$ [14]. Then the 30 meV difference of band splitting between bulk ( 490 meV) and monolayer (460 meV) WSe$_2$ must come from interlayer coupling. That is to say, the interlayer coupling strength in bulk WSe$_2$ is about 30 meV. This is a direct experimental determination for such a microscopic parameter. However, in bulk WSe$_2$, both space inversion symmetry and time inversion symmetry are remained. So the split bands at *K* remain spin-degenerate, that is distinct from the case in monolayer WSe$_2$.

From our measured bands, we can estimate the effective mass at the top of the valence band at *K*. It is found to be −0.*529 m*$_e$ for the upper branch and −0.*532 m*$_e$ for the lower branch of two spin splitting bands. The calculated corresponding values are -0.360 m$_e$ to -0.369 m$_e$, and -0.540 m$_e$ to -0.540 m$_e$, respectively [40, 54]. An earlier ARPES measurement leads to the respective values of -0.35 m$_e$ and -0.49 m$_e$ [29]. As for the effective mass of lower branch band, our result is in very good agreement with these LDA calculations and experimental estimate. While the obtained value from our upper branch band is ~ 50% larger than the aforementioned ones in literatures. Such a big difference could be rationally ascribed to substrate screening effect and local strain of monolayer WSe$_2$ lattice, although the real origin needs to be further investigated. According to our fitting, the effective mass of VBM at *Γ* is 2.344 m$_e$, which is quite close to the band calculation value of 2.582m$_e$ [10] and the previous experimental value of 2.74m$_e$ [27].

## 4  Conclusion:

In conclusion, we have systematically studied the electronic structure of large-area exfoliated monolayer WSe$_2$. Multiple characterizations including optical microscopy, Raman spectroscopy and AFM indicate that the exfoliated WSe$_2$ sample are of excellent homogeneity and high quality within a few millimeter scale. The high-resolution band structures along all high symmetry directions are measured, which provide the direct evidence of direct band-gap semiconductor for such a monolayer $WSe_2$ sample. Surface doping by alkaline metal Rb induces a non-rigid band shift, pointing to importance of electron correlation in engineering its electronic structure. We directly observe the valence-band dispersion around *K* point of monolayer $WSe_2$ and determine its spin splitting of 460 meV, that is in good agreement with typical band calculations and earlier ARPES measurements. Other electronic parameters like valence band effect mass and band gap are also extracted from our ARPES results. In addition, we found that the screening effect from substrate may have substantial effect on these microscopic electronic parameters. Our study provides comprehensive electronic band structure information on monolayer $WSe_2$. It would be useful to deeply understand its electronic properties and develop scalable and high performance electronic devices. By means of our improved mechanical exfoliation method, one may build a valuable and realistic device platform for monolayer $WSe_2$ and other TMDs monolayers in future.

**Electronic Supplementary Material**

This manuscript includes Electronic Supplementary Material as a separate file.

**Notes**

The authors declare no competing financial interest.

**Acknowledgement**

This work is supported by the National Science Foundation of China (11574367, 11874405), the National Key Research and Development Program of China (2016YFA0300600), and the Youth Innovation Promotion Association of CAS

(2017013, 2019007).

## References

[1] Geim, A.K. Graphene: Status and Prospects, *Science,* **2009,** *324,* 1530.

[2] Pacile,D.; Meyer, J. C.; Girit, C.O. and Zett, A. The two-dimensional phase of boron nitride: few-atomic-layer sheets and suspended membranes. *Appl. Phys. Lett.,* **2008,** *92,* 133107

[3] Liu, H.; Neal, A. T.;. Zhu, Z; Luo, Z.; Xu, X.; T´nek, D. and Ye, P. D. Phosphorene: An Unexplored 2D Semiconductor with a High Hole Mobility, *ACS Nano.,* **2014,** *8(4),* 4033

[4] Wang, Q. H.;Zadeh , K. K.; Kis, A.; Coleman, J. N. and Strano, M. S. Electronics and optoelectronics of two-dimensional transition metal dichalcogenides, *Nature Nanotech.,* **2012,** *7,* 699

[5] Wilson,J.A. and Yoffe, A.D. Transition metal dichalcogenides: discussion and interpretation of observed optical, electrical and structural properties. *Adv. Phys.,* **1969,** *18,* 193–335.

[6] Huang, B.; Clark, G.;Moratalla, E. N.; Klein, D. R.; Cheng, R.; Seyler , K. L.; Zhong , D.; Schmidgall , E.; McGuire, M. A.; Cobden , D. H.; Yao, W.; Xiao,D.;Herrero, P. J. and Xu, X. Layer-dependent ferromagnetism in a van der Waals crystal down to the monolayer limit. *Nature ,***2017,** *546,* 270

[7] Deng, Y.; Yu, Y.; Song, Y.; Zhang, J.;Wang, N. Z.; Sun, Z.;Yi, Y.; Wu, Y. Z.; Wu, S.; Zhu, J.; Wang, J.; Chen, X. H. and Zhang, Y. Gate-tunable room-temperature ferromagnetism in two-dimensional $Fe_3GeTe_2$, *Nature,* **2018,** *563,* 94.

[8] Ellis, J.K.; Lucero, M.J. and Scuseria, G.E. The indirect to direct band gap transition in multilayered $MoS_2$ as predicted by screened hybrid density functional theory, *Appl. Phys. Lett.,* **2011,** *99,* 261908.

[9] Mak, K. F.; Lee, C.; Hone, J.; Shan, J. and Heinz, T. F. Atomically Thin $MoS_2$: A New Direct-Gap Semiconductor, *Phys.Rev. Lett.,* **2010,** *105,* 136805 .

[10] Zhu, Z.Y.; Cheng, Y.C and Scwingenshlögl, U. Giant spin-orbit-induced spin splitting in two-dimensional transition-metal dichalcogenide semiconductors, *Phys. Rev. B,* **2011,** *84,* 153402 .

[11] Cheiwchanchamnangij, T.; and Lambrecht, W.R.L. Quasiparticle band structure calculation of monolayer, bilayer, and bulk $MoS_2$, *Phys. Rev. B,* **2012,** *85,* 205302.

[12] Kumar, A. and Ahluwalia, P.K. Electronic structure of transition metal dichalcogenides monolayers 1H-$MX_2$ (M=Mo,W;X=S,Se,Te) from ab-initio theory: new direct band gap semiconductors, *Eur. Phys. J. B,* **2012,** *85,* 186.

[13] Zhang, Y. ; Chang, T. R.; Zhou, B.; Cui, Y. T.; Yan, H.; Liu, Z. K.; Schmitt,  F.; Lee, J.; Moore, R.; Chen, Y. L.; Lin, H.; Jeng, H. T.; Mo, S. K.; Hussain, Z.; Bansil, A. and Shen, Z. X.  Direct observation of the transition from indirect to direct bandgap in atomically thin epitaxial $MoSe_2$, *Nat. Nanotech.,* **2014,** *9,* 111.

[14] Xiao,D.; Liu, G.-B.; Feng, W.; Xu, X. and Yao, W. Coupled Spin and Valley Physics in Monolayers of $MoS_2$ and Other Group-VI Dichalcogenides, *Phys. Rev. Lett.,* **2012,** *108,* 196802.

[15] Cao , T.; Wang, G.; Han, W.; Ye, H.; Zhu, C.; Shi , J.; Niu, Q.;Tan, P.; Wang, E.; Liu, B. and Feng, J. Valley-selective circular dichroism of monolayer molybdenum disulphide, *Nature Commun.,* **2012,** *3,* 887.

[16] Zeng, H.;Dai, J.; Yao, W.; Xiao, D. and Cui, X.  Valley polarization in $MoS_2$ monolayers by optical pumping, *Nature Nanotech.,* **2012,** *7,* 490.

[17] Riley ,J. M.; Mazzola , F.; Dendzik , M.; Michiardi , M.; Takayama, T.; Bawden , L.; Granerød , C.; Leandersson , M.; Balasubramanian , T.; Hoesch , M.; Kim , T. K.; Takagi, H.; Meevasana , W.; Hofmann , Ph.;Bahramy, M. S. ; Wells, J. W. and King, P. D. C.  Direct observation of spin-polarized bulk bands in an inversion-symmetric semiconductor, *Nat. Phys.,* **2014,** *10,* 835

[18] Splendiani, A.; Sun, L.; Zhang, Y.; Li, T.; Kim, T.; Chim, C.-Y.; Galli,G. and Wang, F.  Emerging photoluminescence in monolayer $MoS_2$. *Nano Lett.,* **2010,** *10,* 1271.

[19] Ramasubramaniam,A. Large excitonic effects in monolayers of molybdenum and tungsten dichalcogenides, *Phys. Rev. B,* **2012,** *86,* 115409.

[20] Zhu, B.; Chen, X. and Cui, X. Exciton Binding Energy of Monolayer $WS_2$, *Sci. Rep.,* **2015,** *5,* 9218

[21] Ugeda, M. M.; Bradley, A. J.; Shi, S.-F.;  Jornada, F. H.da;Zhang, Y. ;Qiu, D. Y. ; Ruan, W.; Mo, S.-K.; Hussain, Z.; Shen, Z.-X.;Wang, F. ; Louie,S. G. and Crommie, M. F. Giant bandgap renormalization and excitonic effects in a monolayer transition metal dichalcogenide semiconductor, *Nature Mater.,* **2014,** *13,* 1091.

[22] Radisavljevic, B.; Radenovic, A.; Brivio, J.;Giacometti, V. and Kis, A. Single-layer $MoS_2$ transistors, *Nature Nanotech.,* **2011,** *6,* 147

[23] Chhowalla, M.; Jena, D. and Zhang, H. Two-dimensional semiconductors for transistors, *Nat. Rev. Materials,* **2016,** *1,* 16052

[24] Sanchez,O. L.-; Lembke, D.; Kayci, M.; Radenovic, A. and Kis, A. Ultrasensitive photodetectors based on monolayer $MoS_2$, *Nature Nanotech.,* **2013,** *8,* 497.

[25] Liu,Y.; Weiss,  N. O.; Duan, X.; Cheng, H.-C.; Huang, Y. and Duan, X. Van der Waals heterostructures and devices,*Nat. Rev. Materials,* **2016,** *1,* 16042

[26] Xu, X.; Yao, W.; Xiao, D. and Heinz, T. F. Spin and pseudospins in layered transition metal dichalcogenides, *Nat. Phys.,* **2014,** *10,* 343

[27] Zhang, Y.; Ugeda, M. M.; Jin, C.; Shi, S. F.; Bradley, A. J.; Recio, A. M.;  Ryu,H.; Kim, J. ; Tang, S.; Kim, Y.; Zhou, B.; Hwang,

C.;Chen, Y.; Wang, F.; Crommie, M. F.; Hussain, Z.; Shen, Z. X. and Mo, S.K.  Electronic Structure, Surface Doping, and Optical Response in Epitaxial WSe$_2$ Thin Films, *Nano Lett.,* **2016,** *10,* 2485

[28] Hao, K.; Moody, G.; Wu, F.; Dass, C. K.; Xu, L.; Chen, C.-H.; Sun, L.;  Li,M.-Y.; Li, L.-J.; MacDonald, A. H. and Li, X. Direct measurement of exciton valley coherence in monolayer WSe$_2$, *Nat. Phys.,* **2016,** *12,* 677

[29] Le,D.; Barinov, A.; Preciado, E.; Isarraraz, M.; Tanabe, I.; Komesu, T.; Troha, C.; Bartels, L.; Rahman, T. S and Dowben, P. A. Spin-orbit coupling in the band structure of monolayer WSe$_2$, *J.Phys.:Condens. Matter,* **2015,** *27,* 182201

[30] Wilson,N. R.; Nguyen, P. V.; Seyler, K.; Rivera, P.; Marsden, A. J.;  Laker, Z. P.L.;Constantinescu, G.C.;Kandyba, V. ; Barinov, V. ; Hine, N. D.M.; Xu, X.; Cobden, D. H. Determination of band offsets, hybridization, and exciton binding in 2D semiconductor heterostructures, *Sci. Adv.,* **2017,** *3,* e1601832

[31] Huang,Y.; Sutter, E.; Shi, N. N.; Zheng, J.; Yang, T.; Englund, T.; Gao, H.-J. and Sutter, P. Reliable Exfoliation of Large-Area High-Quality Flakes of Graphene and Other Two-Dimensional Materials, *ACS Nano,* **2015,** *9,* 10612

[32] Liu,G.; Wang, G.; Zhu, Y.; Zhang, H.; Zhang, G.; Wang, X.; Zhou, Y.; Zhang, W.; Liu, H,; Zhao, L.; Meng, J.; Dong, X.;Chen, C.; Xu, Z and Zhou, X. J. Development of a vacuum ultraviolet laser-based angle-resolved photoemission system with a superhigh energy resolution better than 1 meV, *Rev. Sci. Instrum.,* **2008,** *79,* 023105

[33] Coehoorn, R.; Haas, C.; Dijkstra, J.; Flipse, C. J. F. ; Groot, R. A. de; and Wold,A. Electronic structure of MoSe$_2$, MoS$_2$, and WSe$_2$. I. Band-structure calculations and photoelectron spectroscopy，*Phys. Rev. B*，**1987**，*35,* 6195

[34] Blake,P. and Hill, E. W.; Neto, A. H.C.; Novoselov, K. S.; Jiang, D.; Yang, R.; Booth, T. J. and Geim, A. K.  Making graphene visible, *Appl. Phys. Lett.,* **2007,** *91,* 063124

[35] Huang,Y.; Sutter, E.; Sadowski, J. T.; Cotlet, M.; Monti, O. L.A.; Racke, D. A.; Neupane, .M. R.; Wickramaratne, D.; Lake, R. K.; Parkinson,.B. A. and Sutter, P.  Tin Disulfifide;An Emerging Layered Metal Dichalcogenide Semiconductor: Materials Properties and Device Characteristics, *ACS Nano,* **2014,** *8,* 10743

[36] Fang,H.; Chuang, S.; Chang, T.C.; Takei, K.; Takahashi, T.; Javey, A. High-Performance Single Layered WSe$_2$ p-FETs with Chemically Doped Contacts, *Nano Lett.,* **2012,** *12,* 3788.

[37] Liu,W.; Kang, J.; Sarkar, D.; Khatami, Y.; Jena, D.; Banerjee, K. Role of Metal Contacts in Designing High-Performance Monolayer n-Type WSe$_2$ Field Effffect Transistors , *Nano Lett.,* **2013,** *13,* 1983.

[38] Huang,Y. ; Wang, X.; Zhang, X.; Chen, X.; Li, B.; Wang, B.; Huang, M.;  Zhu,C.; Zhang, X.; Bacsa, W. S.; Ding, F. and Ruoff, R. S.  Raman Spectral Band Oscillations in Large Graphene Bubbles, *Phy. Rev. Lett.,* **2018,** *120,* 186104

[39] Zhang,X.; Qiao, X.-F.; Shi, W.; Wu, J.-B.; Jiang, D.-S and Tan, P.-H.  Phonon and Raman scattering of two-dimensional transition metal dichalcogenides from monolayer, multilayer to bulk material, *Chem. Soc. Rev.,* **2015,** *44,* 2757

[40] Zibouche,N.; Kuc, A.; Musfeldt, J. and Heine, T. Transition-metal dichalcogenides for spintronic applications , *Ann. Phys. (Berlin),* **2014,** *526,* No. 9–10, 395

[41] Dendzik,M.; Michiardi, M.; Sanders, C.; Bianchi, C.; Miwa, J. A.; Grønborg, S. S.; Lauritsen, J. V.; Bruix, A.; Hammer, B. and Hofmann, P.  Growth and electronic structure of epitaxial single-layer WS$_2$ on Au(111), *Phy. Rev. B,* **2015,** *92,* 245442

[42] Miwa,J. A.; Ulstrup, S.; Sørensen, S. G.; Dendzik, M.; Čabo, A. G.; Bianchi, M.; Lauritsen, J. V. and Hofmann, P. Electronic Structure of Epitaxial Single-Layer MoS2, *Phy. Rev. Lett.,* **2015,** *114,* 046802

[43] Shanavas, S.K. and Satpathy, S. Effective tight-binding model for MX2 under electric and magnetic fields , *Phy. Rev. B,* **2015,** *91,* 235145

[44] Liu,G.B.; Shan, W.Y.; Yao, Y.; Yao, W. and Xiao, D. Three-band tight-binding model for monolayers of group-VIB transition metal dichalcogenides, *Phy. Rev. B,* **2013,** *88,* 085433

[45] Ci,P. ; Chen,P. ; Kang, J.; Suzuki, R.; Choe, H. S.; Suh, J.; Ko, C.; Park, T.; Shen, K.; Iwasa, Y.; Tongay, S.; Ager, J. W.; Wang, L.-W. and Wu,  J. Quantifying van der Waals Interactions in Layered Transition Metal Dichalcogenides from Pressure-Enhanced Valence Band Splitting, *Nano Lett.* **2017,** *17,* 4982

[46] Dou, X.; Ding, K.; Jiang, D.; Fan, X. and Sun, B.  Probing Spin−Orbit Coupling and Interlayer Coupling in Atomically Thin Molybdenum Disulfide Using Hydrostatic Pressure, *ACS Nano* **2016,** *10,* 1619

[47] Zhang, Y.; Li, H.; Wang, H.; Liu, R.; Zhang, R. and Qiu, Z.-J. On Valence-Band Splitting in Layered MoS$_2$, *ACS Nano,* **2015,** *9,* 8514

[48] Fan, X.; Singh, D. J. and Zheng, W.  Valence Band Splitting on Multilayer MoS$_2$: Mixing of Spin−Orbit Coupling and Interlayer Coupling , *J. Phys. Chem. Lett.* **2016,** *7,* 2175

[49] Alidoust, N.; Bian, G.; Xu, S. Y.; Sankar, R.; Neupane, M.; Liu, C.; Belopolski, I.; Qu, D. X.; Denlinger, J. D.; Chou, F. C.; Hasan, M. Z. , Observation of monolayer valence band spin-orbit effect and induced quantum well states in MoX$_2$. *Nature Communications* **2014,** *5,* 4673.

[50] Smith, N. V. and Fisher, G. B. Photoemission Studies of the Alkali Metals. Il. Rubidium and Cesium, *Phy. Rev. B,* **1971,** *3,* 3662.

[51] Adelung, R.; Brandt, J.; Kipp, L. and Skibowski, M. Reconfiguration of charge density waves by surface nanostructures on TaS$_2$，*Phy. Rev. B,* **2000,** *63,* 165327.

[52] Chiang, T.-C., Photoemission studies of quantum well states in thin films, *Surf. Sci. Rep.,* **2000,** *39,* 181–235.

[53] Absor, M.. A. U. ; Kotaka, H.; Ishii, F. and Saito, M. Strain-controlled spin splitting in the conduction band of monolayer WS$_2$, *Phy. Rev. B,* **2016,** *94,* 115131 (2016)

[54] Korḿanyos, A.; Burkard, G.; Gmitra, M.; Fabian, J.;Zólyomi, V. ;Drummond, N. D.  and Fal'ko, V.  Corrigendum: k.p theory for two-dimensional transition metal dichalcogenide semiconductors, *2D Mater.,* **2015,** *2,* 022001

**Figure 1.** *Crystal structure of 2H-WSe$_2$ bulk and characterization of large-area WSe$_2$ monolayer. (a) Side view of the 2H-WSe$_2$ lattice and unit cell. The yellow and blue balls stand for tungsten and selenium atoms, respectively. (b) Top view of lattice structure for 1H-WSe$_2$. (c) Optical microscope image of 1H-WSe$_2$ monolayer sample. (d) Raman spectra of the sample in (c), acquired at different spots marked as red diamonds. Characteristic modes are labeled for sample WSe$_2$ and substrate Si.*

**Figure 2.** *Overview of electronic structure in monolayer WSe$_2$. (a)-(f) Constant energy contours around K point at different binding energy $E_B$. (g) Brillouin zone and high symmetry points. (h) Original ARPES band structure of monolayer WSe$_2$ (hν=21.2 eV) along high symmetry lines as shown in (g) (blue lines). (i) Second-derivative spectra of raw data in (h). The superimposed dotted lines are from band structure calculation in reference [10] (red) and [40](blue).*

**Figure 3.** *Comparison of valence band splitting between bulk and monolayer WSe$_2$ at K point. (a - b) The measured band splitting in K region along Γ-K-M direction for bulk and monolayer WSe$_2$. (c-d) EDCs of the split bands at K point for the bulk (c) and monolayer (d) sample. The red and orange solid lines are corresponding to the original data, and the blue circles are their Gaussian fitting. The EDC peak positions are marked by black dashed lines. The splitting size is extracted directly as two-peak distance.*

**Figure 4.** *Electronic doping effect of monolayer WSe$_2$ through Rubidium deposition on its surface. Original ARPES band structure along $\Gamma$-K high symmetry direction (cut 1 in panel c) before (a) and after (b) Rb surface doping. (c) Brillouin zone, cuts and high symmetry points. (d) Original band structure after Rb doping measured under a special geometry: along a cut crossing K',marked as cut2 in panel c. Direct band gap is marked with black arrow. (e) EDCs of the valence band at $\Gamma$ point. The red solid line is corresponding to the ARPES spectra (a) of pristine sample. Similarly, the green solid line is for the Rb-doped sample (b). (f) EDCs of the valence band at K point. The curves' definition is the same as in panel (e). The momentum positions for taking EDCs are marked in red (before doping) and green (after doping) dash lines in (a-b). (g) Quantum well state (QWS) near $\Gamma$ after heavy Rb doping. Red dashed line is a guide for QWS band. (h) The Fermi surface after Rb doping. Green lines depict the BZ boundary. Red dashed line is a guide for the QWS-derived Fermi surface.*

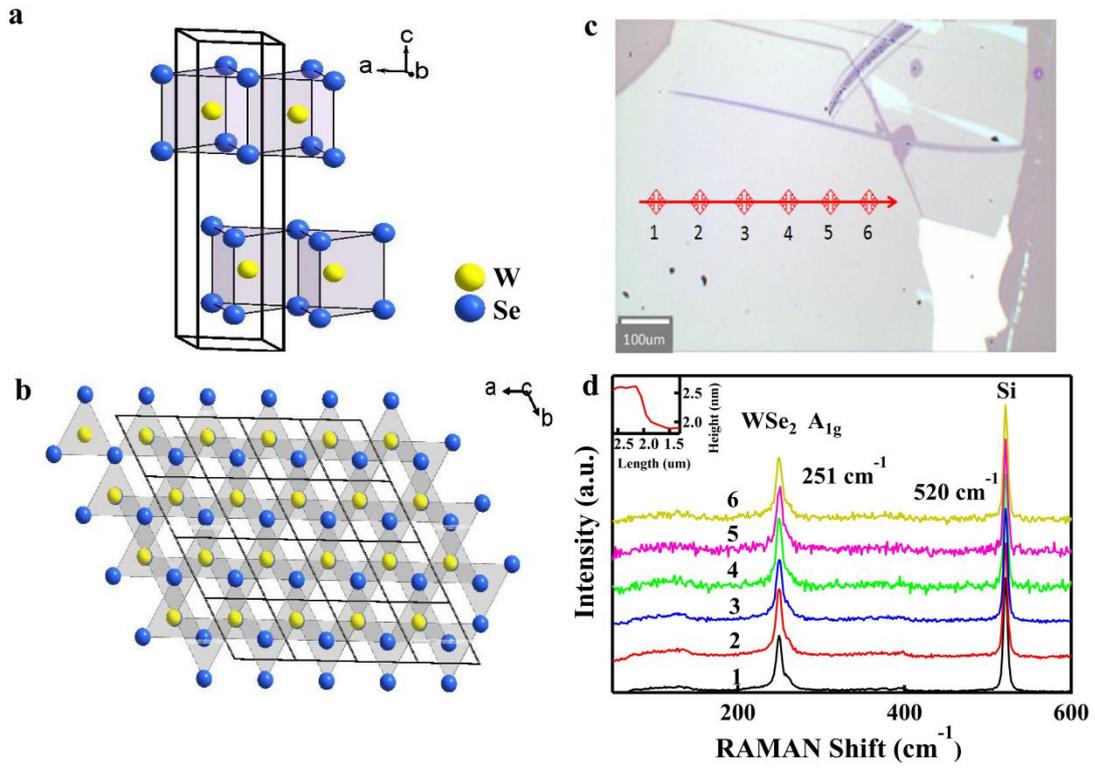

**Figure 1**

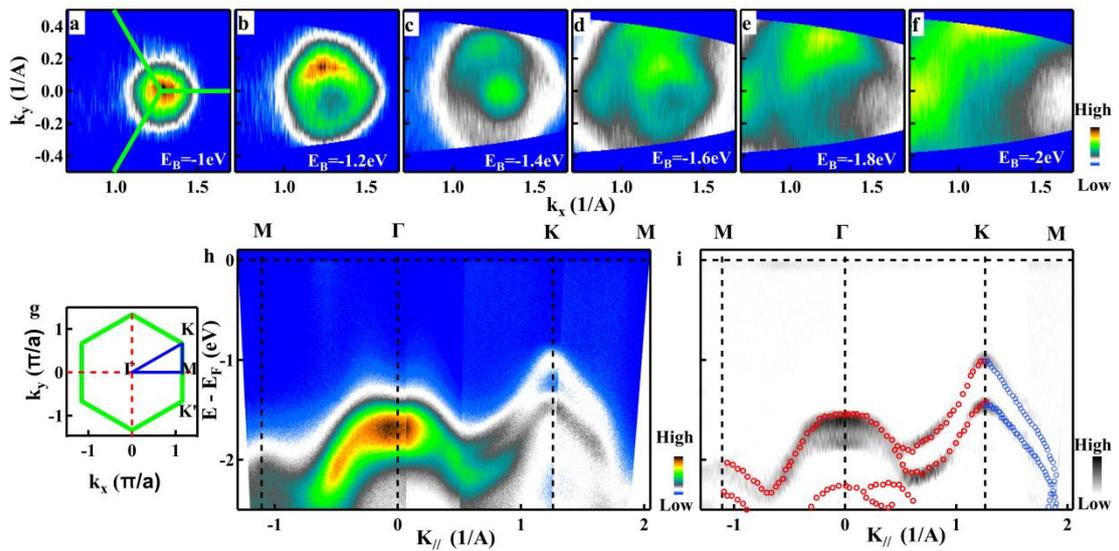

**Figure 2**

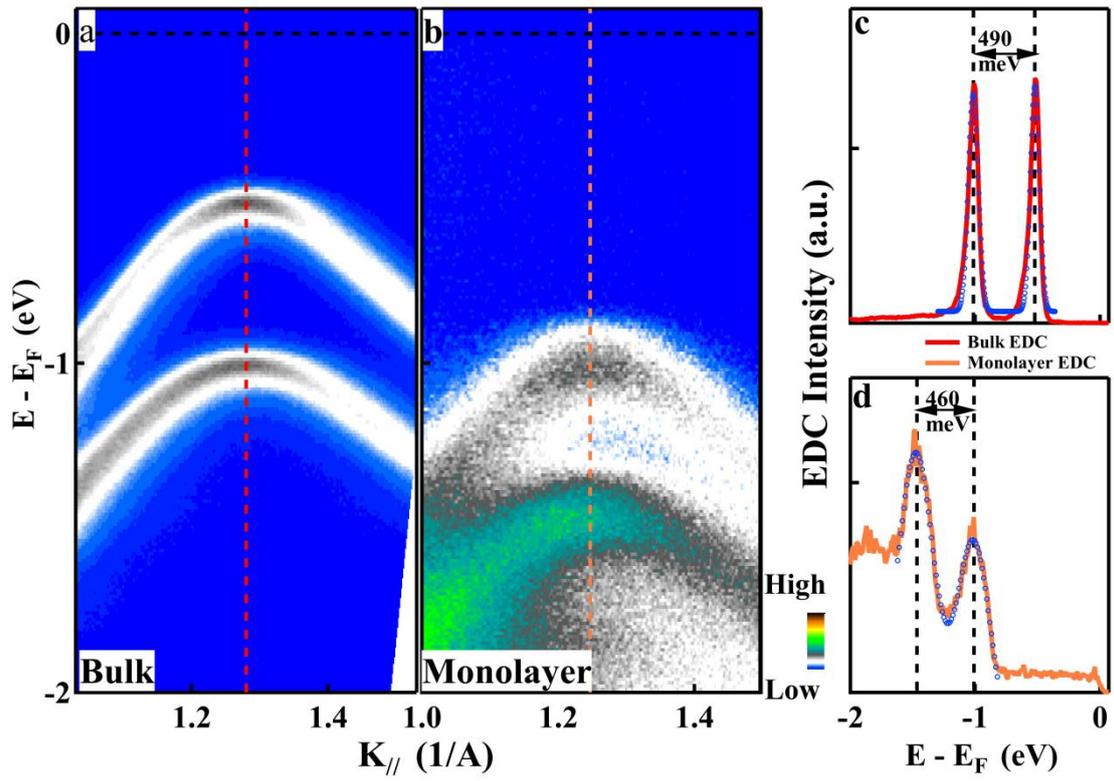

**Figure 3**

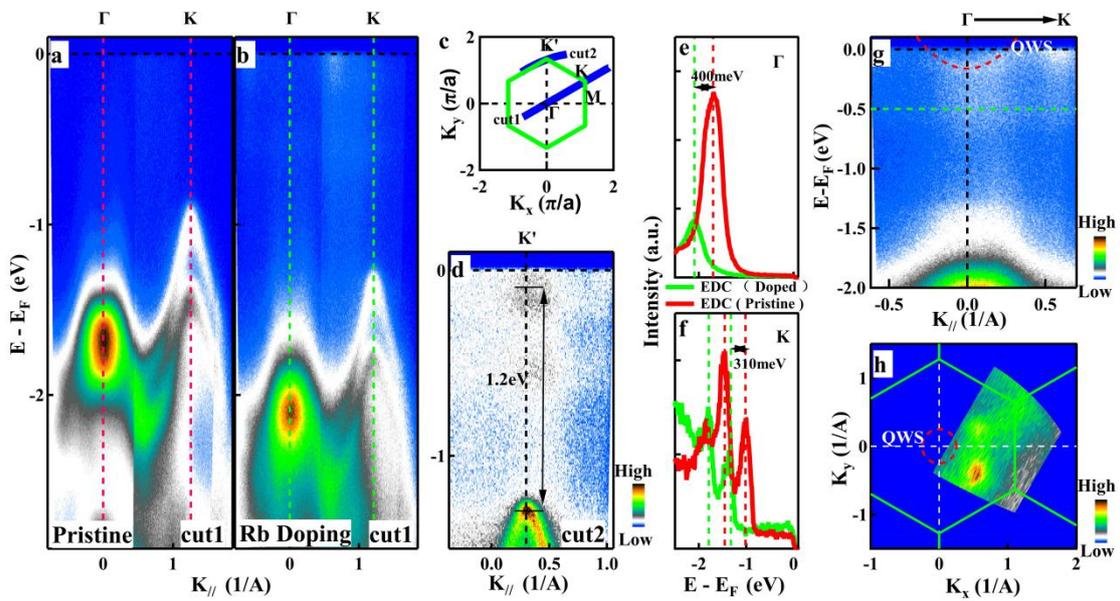

**Figure 4**